\documentclass[referee]{raa}
\usepackage{graphicx,times}
\usepackage{natbib}
\usepackage{amssymb,amsmath}
\usepackage{multirow}
\bibpunct{(}{)}{;}{a}{}{,}

\usepackage[a4paper=true,dvipdfm=true,pagebackref=true]{hyperref}
\hypersetup{pdftitle = The title of my PDF, pdfauthor = My name, pdfsubject= The subject, pdfkeywords = keyword1 keyword2 keyword3}
\hypersetup{colorlinks = true, linkcolor = green, anchorcolor = red, citecolor = blue, filecolor = red, pagecolor = red, urlcolor = red}

\begin{document}

   \title{Portable Adaptive Optics for exoplanet imaging
$^*$
\footnotetext{\small $*$ Supported by the National Natural Science Foundation of China.}
}

 \volnopage{ {\bf 20XX} Vol.\ {\bf X} No. {\bf XX}, 000--000}
   \setcounter{page}{1}

   \author{Yong-Tian Zhu\inst{1,2}, Jiang-Pei Dou$^*$\inst{1,2}, Xi Zhang\inst{1,2}, Gang Zhao\inst{1,2}, Jing Guo\inst{1,2}, Leopoldo Infante\inst{3,4}
   }

   \institute{National Astronomical Observatories / Nanjing Institute of Astronomical Optics \& Technology, Chinese Academy of Sciences, Nanjing 210042, China ; {\it jpdou@niaot.ac.cn}\\
      \and
        Key Laboratory of Astronomical Optics \& Technology, Nanjing Institute of Astronomical Optics \& Technology, Chinese Academy of Sciences, Nanjing 210042, China\\
      \and
        Las Campanas Observatory Carnegie Institution of Washington, Colina El Pino Casilla 601 La Serena, Chile\\
      \and
        Institute of Astrophysics, Pontificia Universidad Cat\'{o}lica de Chile, Av. Vic Mackenna 4860, Santiago, Chile\\
\vs \no
   {\small Received 20XX Month Day; accepted 20XX Month Day}
}

\abstract{The Portable Adaptive Optics (PAO) is a low-cost and compact system, designed for 4-meter class telescopes that have no Adaptive Optics (AO), because of the physical space limitation at the Nasmyth or Cassegrain focus and the historically high cost of the conventional AO. The initial scientific observations of the PAO are focused on the direct imaging of exoplanets and sub-stellar companions. This paper discusses the PAO concept and the associated high-contrast imaging performance in our recent observational runs. PAO is delivering a Strehl ratio better than $60\%$ in $H$ band under median seeing conditions of $1 ''$. Combined with our dedicated image rotation and subtraction (IRS) technique and the optimized IRS (O-IRS) algorithm, the averaged contrast ratio for a $5\leq V\rm_{mag} \leq 9$ primary star is $1.3 \times 10^{-5}$ and $3.3 \times 10^{-6}$ at angular distance of $0.36 ''$ under exposure time of 7 minutes and 2 hours, respectively. PAO has successfully revealed the known exoplanet of $\kappa$ And b, in our recent observation at 3.5-meter ARC telescope at Apache Point Observatory. We have performed the associated astrometry and photometry analysis of the recovered $\kappa$ And b planet, which gives a projected separation of $0.984 \pm 0.05 ''$, a position angle of $51.1 \pm 0.5^{\circ}$, and a mass of $10.15_{-1.255}^{+2.19}M_{\rm Jup}$. These results demonstrate that PAO can be used for direct imaging of exoplanets with medium-sized telescopes.
}

   \authorrunning{Y.-T. Zhu et al. }            
   \titlerunning{Portable Adaptive Optics for exoplanet imaging}  
   \maketitle

%
\section{Introduction}           
  \label{Section_1_Introduction}

With over 3800 already known exoplanets, most of them were discovered by the indirect detection approaches including transit and radial velocity. However, the direct imaging method has become increasingly important, since it can spatially separates the light of a planet from the host star. It has opened a window for the spectroscopy analysis of the atmosphere of the planet, thus will eventually allow the detection of terrestrial biology signals in future space missions. With high-contrast imaging technique, several giant planets have been detected at relatively large orbital separations by ground-based observations \citep{Macintosh2015Science}. The HR 8799 system with four giant planets, for instance, have been observed and recovered mostly by current 8-10 meter class telescopes (\citealp{Marois2008Science}, \citeyear{Marois2010Nature}). The existence of giant planets at large angular separations has challenged the current planet formation model. However, the amount of exoplanets detected by direct imaging is still low comparing with other indirect detection approaches, it therefore requires more high-contrast imaging instruments and the associated scientific observations on as many telescopes as possible, including those small-size telescopes with 2-4 meter aperture.

Most of the discovered exoplanets through direct imaging are, in general, young and self-luminous, and have large orbit separations, which makes them possible to be imaged in infrared and resolved with small telescopes. The observation with small-size telescope has fully demonstrated in Serabyn et al (\citeyear{Serabyn2010}) work. With a 1.5-meter resultant aperture telescope, three of the HR 8799 exoplanets were also recovered by using extreme adaptive optics system that delivers a Strehl ratio of 0.91 at the NIR $K$ band \citep{Serabyn2010}. The high-contrast imaging with small telescopes would be especially important for these time-intensive surveys \citep{Mawet2009}.

We have proposed and built the first PAO that is optimized for exoplanets high-contrast imaging with current 2-4 meter class telescopes since 2013. By employing small-aperture deformable mirror (DM) with micro-electro-mechanical systems (MEMS) technique and the rapid programming of LabVIEW based control software, the PAO features a compact physical size, low cost, high performance and high flexibility that makes it is able to work with any size of telescopes. The PAO with a "small" telescope can provide a high Strehl ratio, since a DM with moderate actuator number can be used and AO high speed correction is achievable.

One of the critical issue for exoplanet high-contrast imaging is the so called non-common path aberrations (NCPA). Both GPI and SPHERE use the two-step approach, which requires a post calibration interferometer or the phase-diversity algorithm to measure the differential aberration then command the DM to correct it (\citealp{Hinkley2009}; \citealp{Fusco2006}; \citealp{Sauvage2007}). In 2012, we presented a unique technique that uses the focal plane PSF information to directly command the DM to correct the NCPA, which was based on an iterative optimization algorithm (\citealp{Dou+2011RAA}; \citealp{Ren2012PASP}). The approach can correct the differential aberration in a single step, which has locked on a wave front with accuracy on the order of $7.4 \times 10^{-4}$ wavelengths.

PAO as a visiting instrument has been used at ESO's 3.6-meter NTT and the 3.5-meter ARC telescope in Apache Point Observatory. In recent observation, PAO has successfully recovered the known exoplanet of $\kappa$ And b under a medium seeing condition. With O-IRS reduction, it has delivered a contrast ratio of $1.3 \times 10^{-5}$ and $3.3 \times 10^{-6}$ at angular distance of $0.36 ''$ under exposure time of 7 minutes and 2 hours, respectively.

In this paper, we present the development and performance of PAO during recent observations. In Section \ref{Section_2_PAOhistory}, we review the development history of PAO; In Section \ref{Section_3_PAOdescription}, we describe the design and associated performance of PAO and provide the observation results especially the analysis of the recovered planet of $\kappa$ And b; conclusions and future work will be presented in Section \ref{Section_4_Conclusions}.

\section{PAO history}
  \label{Section_2_PAOhistory}

PAO is a nighttime AO system that is originated from the Portable Solar Adaptive Optics (PSAO). PSAO is initially proposed for diffraction-limited imaging of the Sun \citep{Ren2009SPIE}. PSAO has been fully tested with different solar telescopes at different seeing conditions, in which it can always be able to deliver good performance (\citealp{Ren2010SPIE}; \citeyear{Ren2012OptEng}; \citeyear{Ren2012PASP}; \citealp{RenAndZhu2013Book}).

We then developed PAO from PSAO by simply replacing the cross-correlation module for daytime wave-front sensing into the centroid-calculation module for nighttime wave-front sensing. The first PAO test observation was conducted in 2013, in which the white dwarf Sirius B in a binary system has been clearly imaged, at NSO's 1.6-meter McMP solar telescope \citep{Ren2014SPIE}.

PAO was then updated and firstly optimized for 4-meter class telescope for more experimental observations since 2014. The PAO prototype system has employed a DM with 97 actuators. With $9\times 9$ wave-front sensor (WFS) sub-apertures, PAO can stably work with a correction speed at 1000 Hz. The first engineering observation was carried out at ESO 3.6-meter NTT telescope in Chile in July, 2014. Six nights was approved by ESO and the observation program was under collaboration with the collaboration of Pontificia Universidad Cat\'{o}lica de Chile, with the support of Chinese Academy of Sciences South America Center for Astronomy (CASSACA). But due to strong wind on the site, only 3 nights were actually allowed to be used during our 6-night observation run, however it was the first time to show our PAO is able to rapidly connect with 4-meter class telescope and provide effective corrections.

\begin{figure}[!h]
  \centering
  \includegraphics[width=10cm]{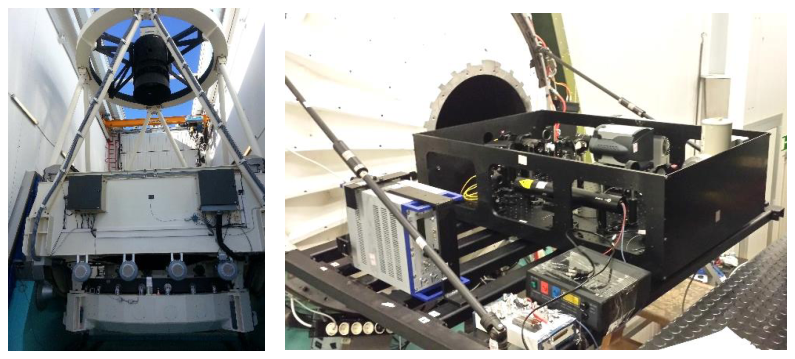}
  \caption{The 3.6-meter NTT telescope and PAO installed on NB.}
  \label{Figure_1_NTTtelescope}
\end{figure}

Figure~\ref{Figure_1_NTTtelescope} shows 3.6-meter NTT telescope (left panel) and our PAO interfaced with the telescope Nasmyth B port (right panel).

\begin{figure}[!h]
  \centering
  \includegraphics[width=10cm]{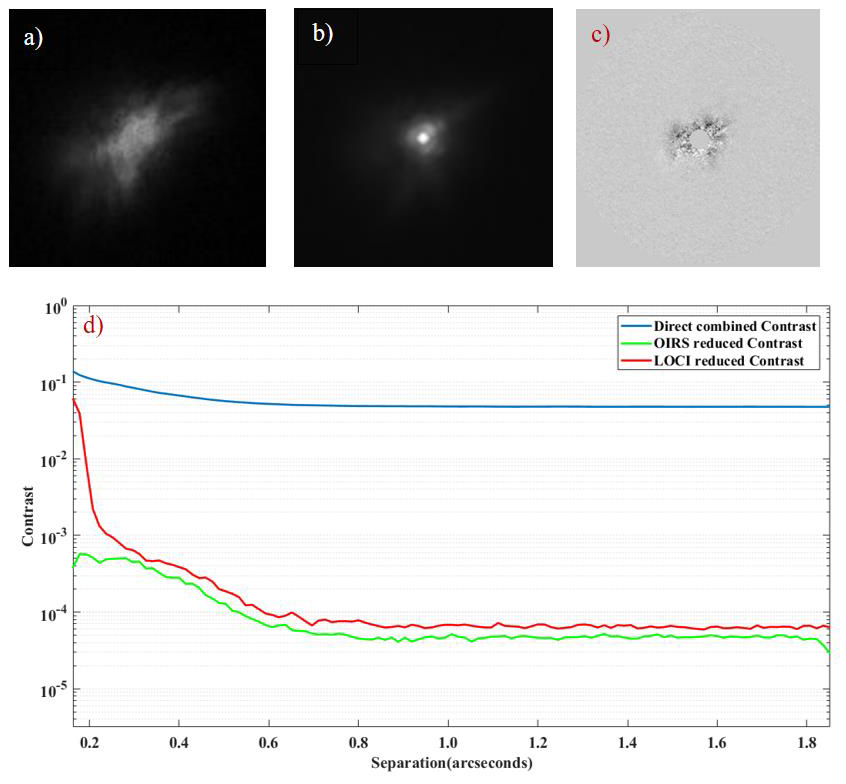}
  \caption{An example of high-contrast reduction for $H$ band observation of Star Fomalhaut.}
  \label{Figure_2_reduction_H_Band}
\end{figure}

Figure \ref{Figure_2_reduction_H_Band} shows an example of a PSF before AO correction (panel a), after AO correction (panel b), and finally processed with O-IRS (panel c). The achieved contrast are shown in panel d). Here we use 144 frames with each frame under a 3 seconds exposure time, which gives a 7 minutes integration time in total. The blue, red and green lines correspond to the direct combination, LOCI \citep{Lafreniere2007} and O-IRS reduction (\citealp{Ren2012ApJ}; \citealp{Dou+2015ApJ}). An initial contrast ratio on the order of $10^{-4}$ has achieved at an angular separation of $0.5 ''$. The above engineering observation clearly demonstrated the feasibility of our PAO technique.

\section{PAO Description}
  \label{Section_3_PAOdescription}

After fully demonstration that the PAO can be used for scientific exoplanet imaging, we purchased more observational time of the 3.5-meter ARC telescope, with the goal to further explore and improve the PAO's capabilities for exoplanet imaging.

\subsection{Optics layout and expected performance}
  \label{Section_3_1_opticalLayout}

\begin{figure}[!h]
  \centering
  \includegraphics[width=10cm]{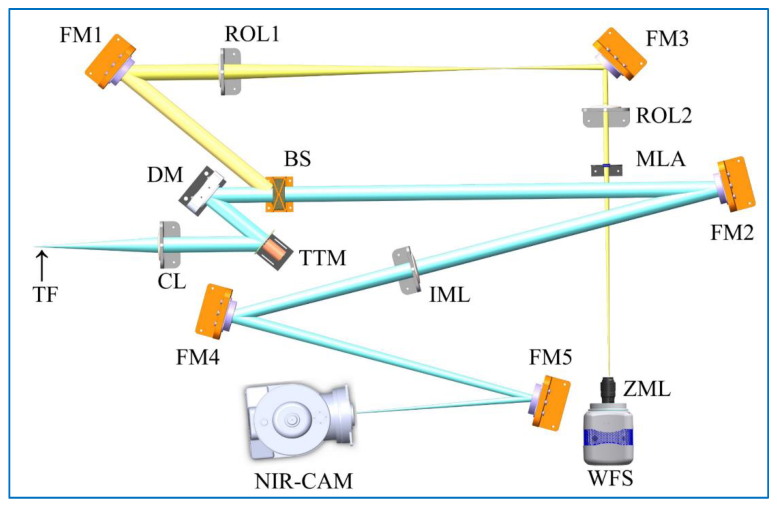}
  \caption{Optics layout of PAO optimized for exoplanet scientific observation.}
  \label{Figure_3_PAOlayout}
\end{figure}

 To further improve the high contrast performance, PAO was finally updated and optimized for the 3.5-meter ARC telescope in Apache Point Observatory located in US. Figure \ref{Figure_3_PAOlayout} shows the optics layout of the PAO for the 3.5-meter ARC telescope. The light from the telescope focus TF (F/\# 10.35) is collimated by the cemented doublet lens with a focal length of 150 mm, in order to match the clear aperture of the DM of 13.5 mm. The tip-tilt mirror (TTM) is located behind the collimator for overall tip-tilt correction. The light propagates to the DM, which corrects possible wave-front errors. A dichroic beam-splitter (BS) is located behind the DM. The visible light is reflected by the beam-splitter to the WFS for wave-front sensing, while the infrared light continuously propagates to a science near infrared (NIR) camera, where the high-contrast image is formed on the focal plane IM at f/50 by the Imaging Lens. The whole PAO system was built on a $0.9 \times 0.78 \times 0.3$ $\rm{m^{3}}$ self-containing enclosed box that will be installed on the ARC Nasmyth A Port.

Employing a high speed 97-element DM and an EMCCD as the WFS camera, PAO could achieve the maximum speed of 1300 Hz for open-loop correction. Table \ref{Table_1_PAOhardware} lists the PAO specifications and estimated performances. The PAO used the DM-97 with a correction speed increased to 2000 Hz, with 10-$\mu$m wave-front stroke. The tip/tilt mirror is mounted on the PI's S-330.4SL piezo tip/tilt platform. The optical deflection angle in the 2 orthogonal axes is 10 mrad and the tilt/tilt platform can achieve a resonance frequency of about 1600 Hz (loaded). The WFS has $9 \times 9$ sub-apertures (excluding those in the 4 corners), each sub-aperture will be sampled by $4 \times 4$ pixels on the WFS camera. The WFS camera is an EMCCD purchased from Photometrics. It has $512 \times 512$ pixels ($16 \times 16$ $\mu$m pixel size), with less than $1 e^{-}$ readout noise in EM gain mode, and can deliver a frame rate of 1820 Hz for our PAO with a region of $64 \times 64$ pixels. The science camera is a NIR camera from Xenics.

\begin{table}[!h]
  \bc
  \begin{minipage}[]{100mm}
  \caption[]{AO Hardware specifications and performance for bright NGSs at good seeing $r_{0}=13$ cm.\label{Table_1_PAOhardware}}\end{minipage}
  \setlength{\tabcolsep}{2.5pt}
  \small
   \begin{tabular}{cc}
    \hline\noalign{\smallskip}
    Hardware & Specifications \\
    \hline\noalign{\smallskip}
    DM       & 97 actuators in $10 \times 10$ configuration, $+/-$ 5-$\mu$m wave-front stroke, \\
             & 13.5-mm clear aperture, 2000 Hz frame rate \\
    TTM      & PI S-330.4SL, 5 mrad stroke range, 1600 Hz frame rate \\
    WFS      & $9 \times 9$ Shack-Hartmann sub-apertures, $4 \times 4$ pixels / sub-aperture \\
    WFS camera          & Photometrics Evolve-512 Delta EMCCD, 1820 Hz at $64 \times 64$ pixels \\
    AO correction speed & 1300 Hz \\
    Computer            & $4 \times 12$-core Intel Xeon$^\circledR$ E5-4657L v2 2.4 GHz (48 cores total) \\
    Field of view       & $30 '' \times 30 ''$ \\
    Strehl ratio at $m_{v}$ = 5  & 0.64 at $H$ \\
    Strehl ratio at $m_{v}$ = 10 & 0.52 at $H$ \\
    \noalign{\smallskip}\hline
  \end{tabular}
  \ec

\end{table}

\begin{figure}[!h]
  \centering
  \includegraphics[width=6cm]{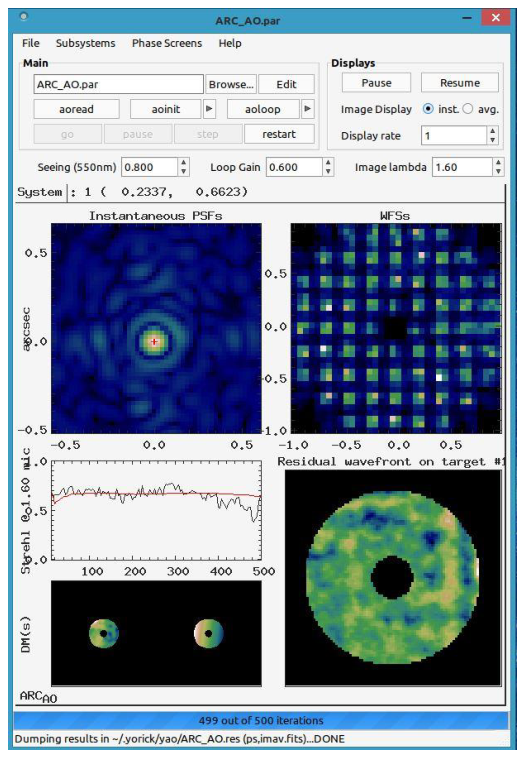}
  \caption{YAO interface for PAO performance simulation.}
  \label{Figure_4_YAOInterface}
\end{figure}

\begin{figure}[!h]
  \centering
  \includegraphics[width=6cm]{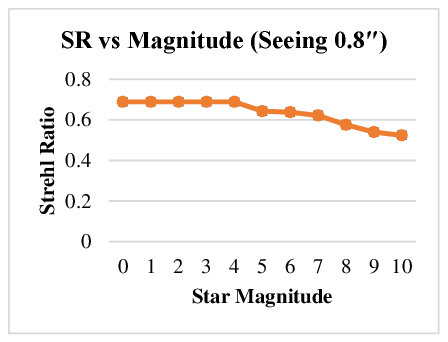}
  \caption{Expected SR correction performance of PAO for different magnitude stars.}
  \label{Figure_5_SRcorrection}
\end{figure}

The error budget of our AO with the SH WFS can be calculated as the residual variance
\begin{equation}
  \label{Eq_residualError}
  \sigma^{2} = \sigma^{2}_{fit} + \sigma^{2}_{r} + \sigma^{2}_{ph} + \sigma^{2}_{bw},
\end{equation}
where $\sigma^{2}_{fit}$ is the fitting error that represents the wave-front variance due to the limited DM actuator number comparing to the telescope aperture, $\sigma^{2}_{r}$ is WFS read noise error, $\sigma^{2}_{ph}$ is the photon noise error, $\sigma^{2}_{bw}$ is the lag error due to finite AO bandwidth, The fitting error is determined by the actuator number which is 97. The read noise is determined by the WFS camera readout noise which is ($0.15 e^{-}$). The photon noise is determined by the NGS magnitude and sub-aperture size. The lag error is determined by the AO correction speed which is 1300 Hz and average wind speed which is 20 m/s. For small residual variance, the Strehl ratio can be calculated by using the analytical equation $S = \exp(-\sigma^{2})$. To precisely evaluate the above errors, it depends on many factors including seeing parameter $r_{0}$, NGS brightness, telescope aperture size and its central obstruction, sub-aperture number of SH WFS, as well as the telescope and SH WFS/DM geometric configurations and other properties, which refer to complicate analytical process. Here we calculate the Strehl ratios (SR) using the end-to-end software YAO simulation software package (http://frigaut.github.io/yao/), with all above error sources and configurations inputted (fitting error, read noise, photon noise, bandwidth, telescope and WFS/DM geometric configurations etc.), and this yields more realistic results. Figure \ref{Figure_4_YAOInterface} shows the interface of the YAO software for the simulation of PAO correction of bright stars. Since the non-common-path error will be corrected by our SPGD algorithm, it is not included in the above calculation/simulation. The AO performance is a function of the seeing condition as well as the guide star brightness. At the good seeing $r_{0}=13$ cm at 500-nm, the Strehl ratio is 0.64 at the $H$ (1.60-$\mu$m) band, for a natural guide star with $m_{v}=5$, as shown in Table \ref{Table_1_PAOhardware} and Figure \ref{Figure_5_SRcorrection}. The AO should deliver better performances at excellent seeing conditions.

\begin{figure}[!h]
  \vspace{1em}
  \centering
  \includegraphics[width=10cm]{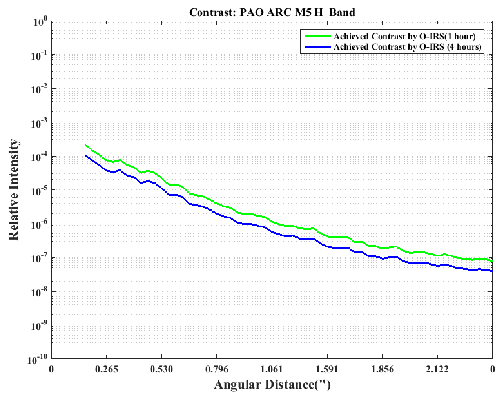}
  \caption{Expected contrast performance of PAO for $M_{v} = 5$ star at $H$ band.}
  \label{Figure_6_PAO_mv5_Hband}
\end{figure}

The simulation software YAO can calculate the PAO instant PSF. The AO corrected PSFs are variable with a period of speckle life $t_{0}$, which is determined as $ t_{0} = 0.6 D/v $ \citep{Macintosh2005SPIE}, where $ D $ is telescope aperture diameter and $v$ is the atmosphere wind speed. For an AO system with the NCPA corrected, the contrast can be improved by increasing the exposure time, with improving factor as $\sqrt{T/t_{0}}$, where $T$ is the exposure time. Figure \ref{Figure_6_PAO_mv5_Hband} shows the expected achievable contrast for the $H$ band exoplanet imaging with $m_{v} = 5$ NGS with an exposure time of 1 hour and 4 hours, respectively. This is achieved by using our unique O-IRS reduced of all instant PSFs, each with an exposure of speckle life \citep{Dou+2015ApJ}. The simulation results prove that the PAO can deliver a contrast $10^{-5.7}$ and $10^{-6}$ at an angular separation of $1''$, for an exposure of 1 hour and 4 hours respectively.

\subsection{PAO Performance at 3.5-meter ARC Telescope}
  \label{Section_3_2_PAOperformance}

\subsubsection{Observation of the binary stars HR 6212 A and B}
  \label{Section_3_2_1_ARCobservation}

\begin{figure}[!h]
  \centering
  \includegraphics[width=10cm]{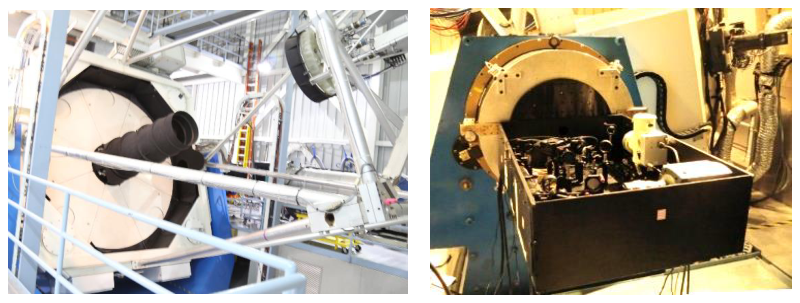}
  \caption{The 3.5-meter ARC telescope and the PAO installed on NA2.}
  \label{Figure_7_ARCtelescope}
\end{figure}

Six observation runs have been conducted at 3.5-meter ARC telescope in Apache Point Observatory during 2015 to 2018. Figure \ref{Figure_7_ARCtelescope} shows the 3.5-meter ARC telescope (left panel) and our PAO interfaced with the telescope Nasmyth port (right panel).

\begin{figure}[!h]
  \centering
  \includegraphics[width=10cm]{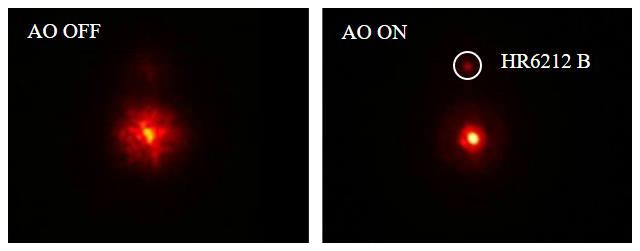}
  \caption{HR 6212 A/B images before and after PAO correction at 3.5-meter ARC telescope.}
  \label{Figure_8_ARC_HR6212AB}
\end{figure}

\begin{figure}[!h]
  \centering
  \includegraphics[width=10cm]{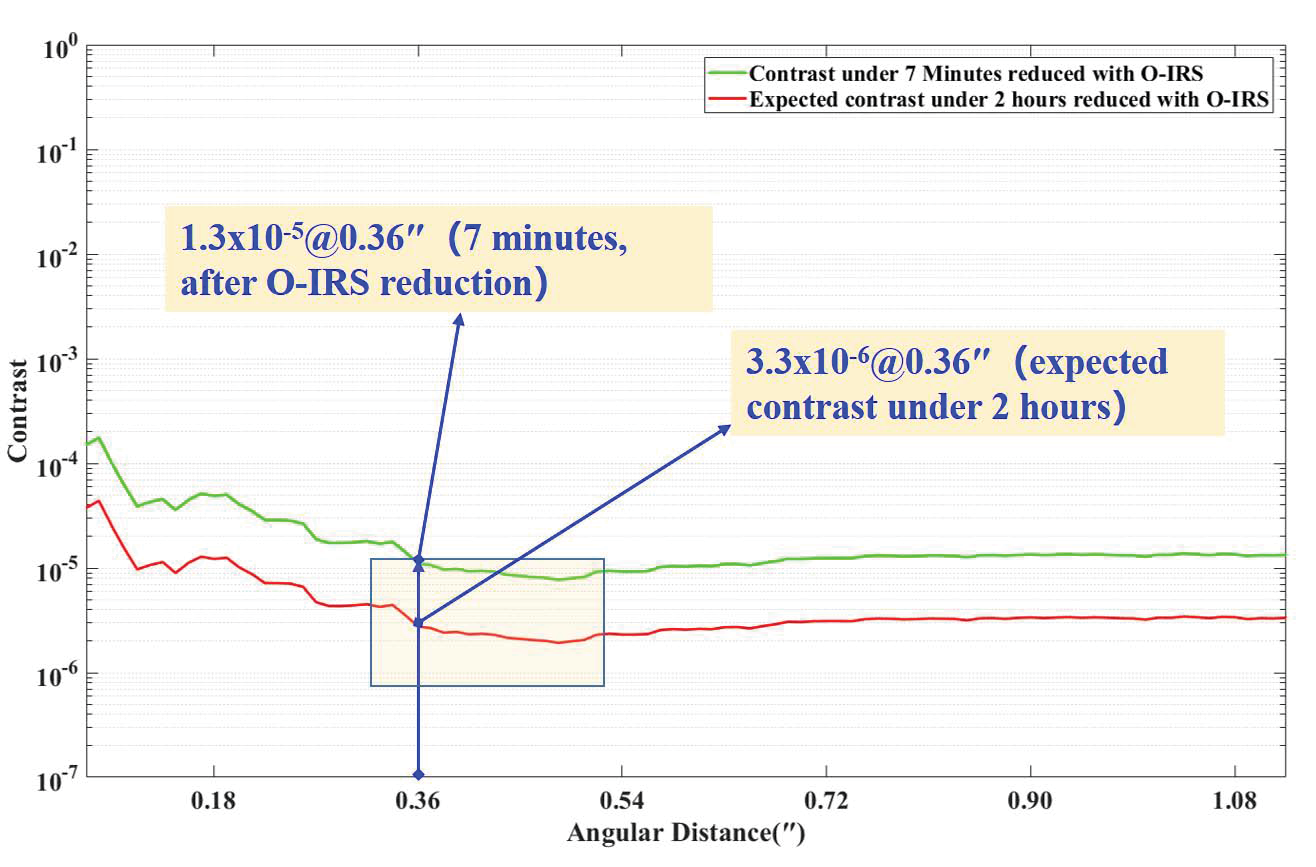}
  \caption{Contrast achieved with PAO at 3.5-meter ARC telescope.}
  \label{Figure_9_ARCcontrast}
\end{figure}

The first two runs was carried out in 2015. A couple of targets were used to test PAO potential performance for high-contrast imaging. Figure \ref{Figure_8_ARC_HR6212AB} shows the AO on and off correction of the binary stars HR 6212 A and B, in which the companion star B was clearly shown after PAO correction. Figure \ref{Figure_9_ARCcontrast} shows the associated contrast achieved with the APO 3.5-meter telescope with an exposure of 7 minutes after O-IRS data reduction. A contrast of $1.3 \times 10^{-5} $ was achieved at an inner working angle (IWA) of $0.36''$, corresponding to $4 \lambda / D$, four times of the coronagraph diffraction beam, which implies that the contrast of $3.3 \times 10^{-6} $ should be achievable with a longer exposure time of 2 hours. The Extreme AOs in routine operation such as GPI and SPHERE on 8-meter telescopes has better performance than current PAO, which can deliver better imaging contrasts of $10^{-6}$ from the angular distance of $0.3''$, considering the Extreme AOs use denser sub-apertures to sample the telescope pupil. However the PAO still showed its capability of imaging exoplanets and brown dwarfs on middle-sized telescopes. The data reduction was conducted by using our O-IRS algorithm \citep{Dou+2015ApJ}. This result clearly indicates that our portable adaptive optics is a mature technique and is ready to be used for scientific exoplanet imaging.

\subsubsection{Observation of the known exoplanet $\kappa$ And b}
  \label{Section_3_2_2_kappaAndbObs}

We firstly conducted the high-contrast imaging observation of $\kappa$ And b in 2014 in $H$-band at 3.5-meter ARC in APO Observatory to testify PAO's capability of direct imaging of giant exoplanets around young stars. The planet $\kappa$ And b was firstly imaged in 2012 by using the Subaru/HiCIAO during the SEEDS survey \citep{Carson2013} and then characterized by the subsequent New Keck and LBTI high-contrast observations in 2012 and 2013 \citep{Bonnefoy2014}. These observations were carried out in $J$, $H$, $Ks$ and $Lp$ infrared bands respectively. For our current NIR camera is cut-off at 1.55 $\mu$m, in this observation run, we only take the $H$ band data for demonstration purpose.

\begin{figure}[!h]
  \centering
  \includegraphics[width=6cm]{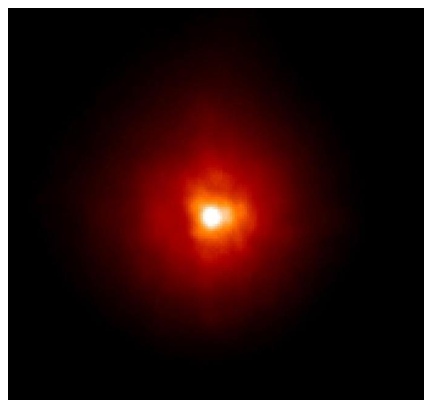}
  \caption{The observed $\kappa$ And image after PAO correction under seeing of $1 ''$.}
  \label{Figure_10_PAO_kappaAndB}
\end{figure}

The observation was carried out with the NIR camera using the $ 640 \times 512 $ pixels array and under the angular differential imaging (ADI) mode without any field rotation compensation. All frames of the host star $\kappa$ And were acquired under 2 seconds exposure time without saturation for the following photometric measurement to derive the flux ratios, in order to compare this observation result with the previous ones. The data were saved in a cube with 50 frames in each cube and 23 cubes in total. The integration time of all cubes were 2300 s with a small field rotation of 8.83 degrees. Figure \ref{Figure_10_PAO_kappaAndB} shows one single frame image of $\kappa$ And after PAO correction.

\subsection{Data reduction}
  \label{Section_3_3_dataReduction}

The raw images were processed based on our recently developed data reduction pipeline for high contrast imaging \citep{Dou+2015ApJ}. We firstly carried out the basic reduction including bad pixels and cosmic rays, dark current, flat field and background calibrations; then these images were registered and conducted the speckle suppression based on O-IRS and PCA, respectively (\citealp{Ren2012ApJ}; \citealp{Dou+2015ApJ}; \citealp{Amara2012}; \citealp{Otten2017}). The pipeline is written in both IDL and Python, a well-known and frequently used language in astronomical environments.

\begin{figure}[!h]
  \centering
  \includegraphics[width=6cm]{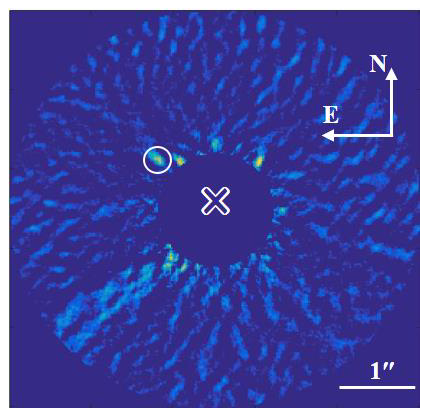}
  \caption{The reduced image of planet $\kappa$ And b, the star center is marked with cross.}
  \label{Figure_11_reducedKappaAndb}
\end{figure}

Meanwhile, we performed the astrometry analysis of the planet b, in which we took into account the positional uncertainties of the centroid errors of the companion, and systematic errors in the distortion solution and the north/west alignment. For the orientation alignment, we have recorded two cubes of images when moving the telescope toward north (distance of $2.7 ''$) firstly then to the east (distance of $2.9 ''$), respectively. The image rotation angle was calculated by using the center of the median of all images in the two cubes, which gives a rotation angle of $4.5 \pm 0.5^{\circ}$ to align the north orientation with the y axis of the image. And the pixel scale of PAO image is found out to be $0.024 ''$/pixel. Finally, the reduced image of $\kappa$ And b processed by the PCA-based algorithm was de-rotated of $4.5^{\circ}$ and shown in Figure \ref{Figure_11_reducedKappaAndb}. Table 2 listed the astrometry analysis results based on previous observation in 2012-2013 and in this observation.

\begin{table}[!h]
  \bc
  \begin{minipage}[]{100mm}
  \caption[]{Astrometry analysis of planet $\kappa$ And b.\label{Table_2_astrometryAnalysis}}\end{minipage}
  \setlength{\tabcolsep}{2.5pt}
  \small
   \begin{tabular}{ccccccc}
    \hline\noalign{\smallskip}
    \multirow{3}*{Date} & \multicolumn{2}{c}{$H$ band}          & \multicolumn{2}{c}{$Ks$}              & \multicolumn{2}{c}{$H$ band}          \\
                        & \multicolumn{2}{c}{(1.63 $\mu$m)}     & \multicolumn{2}{c}{(2.65 $\mu$m)}     & \multicolumn{2}{c}{(1.55 $\mu$m)}     \\
                        & Proj. sep. ($''$) & Proj. ang. ($''$) & Proj. sep. ($''$) & Proj. ang. ($''$) & Proj. sep. ($''$) & Proj. ang. ($''$) \\
    \hline\noalign{\smallskip}
    2012 January 1  & $1.070 \pm 0.010$ & $55.7 \pm 0.6$ & & & &  \\
    2012 July 8     & $1.058 \pm 0.007$ & $56.0 \pm 0.4$ & & & &  \\
    2012 October 30 & & & \multirow{2}*{$1.029\pm0.005$} & \multirow{2}*{$55.3\pm0.3$} & & \\
    2012 November 3 & & &                                &                             & & \\
    2017 June 10    & & & & & $0.984\pm0.05$ & $51.1\pm0.5$ \\
    \noalign{\smallskip}\hline
    Reference       & \multicolumn{2}{c}{(1)} & \multicolumn{2}{c}{(2)}&\multicolumn{2}{c}{(3)} \\
    \noalign{\smallskip}\hline
  \end{tabular}
  \ec
  \tablecomments{0.86\textwidth}{References: (1) \citealp{Carson2013}; (2) \citealp{Bonnefoy2014}; (3) this work.}
\end{table}

\begin{figure}[!h]
  \centering
  \includegraphics[width=6cm]{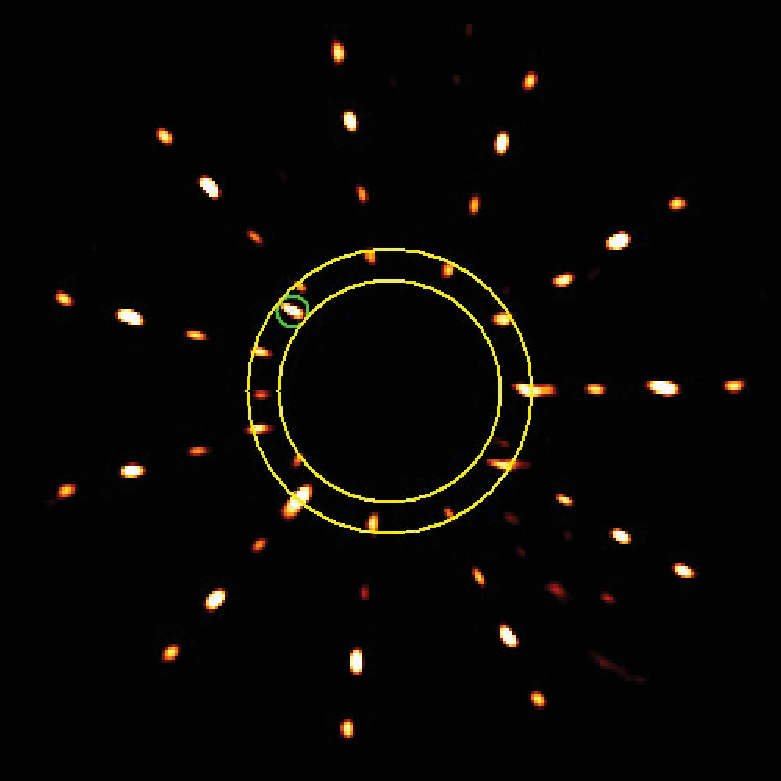}
  \caption{The reduced image of $\kappa$ And with artificial planets added: the location of planets b is marked in the green circle, while the yellow annulus shows the second innermost artificial planets with the same angular separation of planet b.}
  \label{Figure_12_reducedKappa}
\end{figure}

We then performed the photometry analysis of the reduced planets. A series of artificial planets were added in the same data of $\kappa$ And after basic reduction with a certain contrast, with a similar procedure that has been used in previous work (\citealp{Lafreniere2007}; \citealp{Dou+2015ApJ}). In each frame of the data set, planets are inserted at eleven different azimuth angles. In order to cover the region where planet b is located, here several artificial planets along one radius were added, in which the second innermost one has an angular separation of $0.984''$, the same as plant b. The final frame has a field rotation of 8.83 degrees, which is also the same as the data set. Then the speckle suppression based on O-IRS and PCA are carried on these images in the data set, respectively. Figure \ref{Figure_12_reducedKappa} shows the reduced image with artificial planets by using the PCA-based algorithms, when adjusting the intensity of the artificial planet to be the same as the planet b. It finally gives a contrast of $4.691 \pm 0.85 \times 10^{-5}$, which corresponding to the magnitude of $15.145 \pm 0.195$ at 1.55 $\mu$m. In Figure \ref{Figure_12_reducedKappa}, an annulus between $0.895''$ and $1.072''$ has added to cover the location of planet $\kappa$ And b. The signal to noise ratio (S/N) of the planet has been calculated, which is 7.75 $\sigma$.

A summary of the available photometry is presented in Table 3. Here we employ the same physical parameters of the primary star of $\kappa$ And b as the work in Carson et al. 2013, in which the estimated age is $30_{-10}^{+20}$ Myr, and the parallax is $19.2 \pm 0.7$ \citep{Perryman1997}. The $H$-band (1.55 $\mu$m) magnitude is $15.145 \pm 0.195$ gives a mass of $10.15^{+2.19}_{-1.255}M_{\rm{Jup}}$ and a temperature of $1227_{-4.5}^{+4}$ K by employing the BT-DUSTY evolutionary models (\citealp{Chabrier2000}; \citealp{Spiegel2012}).

\begin{table}[!h]
  \bc
  \begin{minipage}[]{100mm}
  \caption[]{Photometry analysis of planet $\kappa$ And b.\label{Table_3_photometryAnalysis}}\end{minipage}
  \setlength{\tabcolsep}{2.5pt}
  \small
   \begin{tabular}{cccccccc}
    \hline\noalign{\smallskip}
    \multirow{2}*{Planets} & $J$ band     & $H$ band    & $H$ band     & $Ks$         & $L'$        & $NB$         & $M$ \\
                           & (1.25$\mu$m) & (1.63$\mu$m) & (1.55$\mu$m) & (2.65$\mu$m) & (3.8$\mu$m) & (4.05$\mu$m) & (4.7$\mu$m) \\
    \hline\noalign{\smallskip}
    \multirow{2}*{$\kappa$ And b} & $15.86\pm0.21$ & $14.95\pm0.13$ & $15.145\pm0.195$ & $14.32\pm0.09$ & $13.12\pm0.1$ & $13.0\pm0.2$ & $13.3\pm0.3$ \\
                                  & $16.3 \pm 0.3$ & $15.2 \pm 0.2$ &                  & $14.6 \pm 0.4$ & $13.12\pm0.09$&              &              \\
    \hline\noalign{\smallskip}
    Reference                     & (1), (2)       & (1), (2)       & (3)              & (1), (2)         & (1), (2)    & (1)          & (1) \\
    \noalign{\smallskip}\hline
  \end{tabular}
  \ec
  \tablecomments{0.86\textwidth}{References: (1) \citealp{Bonnefoy2014}; (2) \citealp{Carson2013}; (3) this work.}
\end{table}

\section{Conclusions}
  \label{Section_4_Conclusions}

In this paper, we have briefly reviewed the development history of PAO technique. We then update the PAO to be a visiting instrument that is optimized for the direct imaging of exoplanets and sub-stellar companions. The unique feature of PAO with a compact physical size and high performance allows it to be used at current 3-4 meter class telescopes. In the most recent observation at 3.5-meter ARC telescope at APO, we have successfully recovered the known exoplanet of $\kappa$ And b reduced by using our unique O-IRS and PCA algorithm, respectively. The observation contrast has reached $10^{-5}$ at an angular distance from $0.36 ''$ to $1 ''$. Finally, we have performed the associated astrometry and photometry analysis of the recovered $\kappa$ And b planet, which is consistent with the previous work. It has fully demonstrated that our portable high-contrast imaging system can be used for the exoplanets direct imaging observation with middle-class telescopes.

\normalem
\begin{acknowledgements}

Y.-T. Zhu acknowledges the support from the National Natural Science Foundation of China (NSFC) under grant \# 11827804.
J.-P. Dou acknowledges the support from the National Natural Science Foundation of China (NSFC) under grant \# U2031210.
\end{acknowledgements}

\bibliographystyle{raa}
\bibliography{PAOreference}

\end{document}